\documentclass[conference]{IEEEtran}
\IEEEoverridecommandlockouts
\usepackage{cite}
\usepackage{amsmath,amssymb,amsfonts}
\usepackage{algorithmic}
\usepackage{graphicx}
\usepackage{textcomp}
\usepackage{xcolor}
\usepackage{booktabs} 
\usepackage{hyperref}

\renewcommand{\figureautorefname}{Figure~\negthinspace}

\renewcommand{\sectionautorefname}{Section~\negthinspace}

\def\BibTeX{{\rm B\kern-.05em{\sc i\kern-.025em b}\kern-.08em
    T\kern-.1667em\lower.7ex\hbox{E}\kern-.125emX}}
\begin{document}

\title{Toward Large-Scale Distributed\\ Quantum Long Short-Term Memory\\ with Modular Quantum Computers

\thanks{The views expressed in this article are those of the authors and do not represent the views of Wells Fargo. This article is for informational purposes only. Nothing contained in this article should be construed as investment advice. Wells Fargo makes no express or implied warranties and expressly disclaims all legal, tax, and accounting implications related to this article. \IEEEauthorrefmark{1} Corresponding Author: kuan-cheng.chen17@imperial.ac.uk}
}

\author{
\IEEEauthorblockN{
    Kuan-Cheng Chen\IEEEauthorrefmark{2}\IEEEauthorrefmark{3}\IEEEauthorrefmark{1},
    Samuel Yen-Chi Chen\IEEEauthorrefmark{4}, 
    Chen-Yu Liu\IEEEauthorrefmark{5}, 
    Kin K. Leung\IEEEauthorrefmark{2}
}
\IEEEauthorblockA{\IEEEauthorrefmark{2}Department of Electrical and Electronic Engineering, Imperial College London, London, UK}
\IEEEauthorblockA{\IEEEauthorrefmark{3}Centre for Quantum Engineering, Science and Technology (QuEST), Imperial College London, London, UK}
\IEEEauthorblockA{\IEEEauthorrefmark{4}Wells Fargo, New York, NY, USA}
\IEEEauthorblockA{\IEEEauthorrefmark{5}Graduate Institute of Applied Physics, National Taiwan University, Taipei, Taiwan}
}

\maketitle

\begin{abstract}
In this work, we introduce a Distributed Quantum Long Short-Term Memory (QLSTM) framework that leverages modular quantum computing to address scalability challenges on Noisy Intermediate-Scale Quantum (NISQ) devices. By embedding variational quantum circuits into LSTM cells, the QLSTM captures long-range temporal dependencies, while a distributed architecture partitions the underlying Variational Quantum Circuits (VQCs) into smaller, manageable subcircuits that can be executed on a network of quantum processing units. We assess the proposed framework using nontrivial benchmark problems such as damped harmonic oscillators and Nonlinear Autoregressive Moving Average sequences. Our results demonstrate that the distributed QLSTM achieves stable convergence and improved training dynamics compared to classical approaches. This work underscores the potential of modular, distributed quantum computing architectures for large-scale sequence modeling, providing a foundation for the future integration of hybrid quantum-classical solutions into advanced Quantum High-performance computing (HPC) ecosystems.
\end{abstract}


\section{\uppercase{Introduction}}
\label{sec:introduction}
The exponential success of machine learning (ML), especially deep learning (DL), in fields such as image recognition, natural language understanding, and complex decision-making has led to increasingly sophisticated frameworks that capture intricate data structures\cite{lecun2015deep}. Among these frameworks, recurrent neural networks (RNNs) play a pivotal role in modeling temporal or sequential information\cite{medsker2001recurrent,hochreiter1997long}. Their ability to represent dynamic, time-dependent patterns has made them central to a range of applications, from classical tasks like speech recognition \cite{graves2013speech}, machine translation \cite{sutskever2014sequence} and financial forecasting\cite{cao2019financial} to advanced physics problems involving chaotic dynamics \cite{mikhaeil2022difficulty}. Notably, recent studies have explored how RNNs can approximate and control the behavior of quantum systems, indicating that temporal modeling techniques can be extended into the domain of quantum information processing \cite{flurin2020using}.

On the hardware front, the past few years have witnessed a surge in the development and commercialization of quantum processors\cite{ladd2010quantum,de2021materials}. These noisy intermediate-scale quantum (NISQ) devices \cite{preskill2018quantum} are not yet equipped with full-scale quantum error correction but can still offer unique computational advantages for certain classes of problems \cite{huang2022quantum}. Variational quantum circuits (VQCs) have emerged as a promising paradigm to leverage NISQ hardware \cite{bharti2022noisy}: They use classical optimization loops to train parameterized quantum gates, effectively absorbing hardware noise and capitalizing on quantum resources like entanglement. Consequently, VQCs have demonstrated utility in classical ML analogues such as classification \cite{mitarai2018quantum,schuld2020circuit,chen2022quantumCNN,hsu2024quantum}, generative modeling \cite{chu2023iqgan}, and reinforcement learning \cite{chen2020variational,lockwood2020reinforcement,chen2024quantum,yun2023quantum,skolik2021quantum,jerbi2021variational,chen2024validating}, hinting at opportunities to incorporate quantum-enhanced approaches into more complex learning architectures.

While previous efforts have established foundational links between VQCs and ML tasks, the challenge of modeling long-term temporal dependencies in quantum systems has remained largely unexplored. To bridge this gap, we focus on long short-term memory (LSTM) networks \cite{hochreiter1997long}, a key variant of RNNs that can capture extended temporal correlations. The LSTM mechanism can integrate naturally with VQCs, yielding a hybrid quantum-classical architecture—referred to as the quantum LSTM (QLSTM)—capable of learning sequential patterns in a way that might surpass classical approaches \cite{chen2022quantumLSTM,lin2024quantum,hsu2024qklstm}. By encoding input signals into quantum states and utilizing quantum operations, quantum-enhanced method can potentially exploit entanglement and other quantum properties to achieve more efficient learning of complex temporal structures than its classical counterpart\cite{chehimi2024federated,cao2023linear,lin2024quantum,hsu2024quantum}. Furthermore, quantum-enhanced machine learning algorithms can offer better expressivity across various tasks, including image processing\cite{senokosov2024quantum,chen2024compressedmediq}, signal processing\cite{lin2024quantum,chen2024consensus}, high-energy physics\cite{guan2021quantum}, climate change\cite{ho2024quantum} and natural language processing\cite{yang2022bert,di2022dawn,stein2023applying,liu2024quantum}, and provide advantages in model compression\cite{liu2024quantum}. 


An innovation of this work is the demonstration that QLSTM approach can be scaled to larger problem instances through a modular, distributed quantum computing paradigm\cite{cuomo2020towards,zhang2019modular,monroe2014large}. Rather than relying on a single, large quantum circuit, we decompose the underlying VQCs into smaller subcircuits, each of which can be executed on a separate quantum processing unit (QPU). In this distributed setting, multiple quantum cores collectively implement a larger quantum model, effectively creating a multi-core quantum computing infrastructure\cite{escofet2024revisiting}. This strategy enhances scalability and facilitates the integration of QLSTM models into large-scale quantum high-performance computing (HPC) ecosystems. By implementing a modular VQC-based architecture, we establish a proof-of-concept that handling complex temporal modeling tasks—such as those involving non-Markovian dynamics in open quantum systems—can become more feasible on near-term quantum devices, potentially accelerating progress toward scalable quantum algorithms.


The rest of the paper is organized as follows: In \sectionautorefname{\ref{Sec.II}}, we review the principles of RNNs and LSTMs in the classical ML context. Then we introduce VQCs as the fundamental building block of our hybrid architectures. We detail the QLSTM formulation and our distributed framework for multi-core QPU integration. In \sectionautorefname{\ref{Sec.III}} , we present numerical results on a variety of temporal tasks, including challenging open quantum system scenarios, and benchmark distributed QLSTM against centric QLSTM performance. 

\section{Methodology}
\label{Sec.II}

In this section, we present a cohesive methodological framework starting from the fundamental principles of classical RNNs and LSTM architectures, and then generalizing these concepts to their quantum counterparts, namely the QLSTM. We further detail how to scale the QLSTM to a distributed form by partitioning its computations across multiple QPUs.

\subsection{Classical RNNs and LSTM Networks}

\subsubsection{Recurrent Neural Networks}

RNNs are a class of neural architectures designed to process sequential data. By introducing feedback connections, RNNs can incorporate information from previous time steps, thereby modeling temporal correlations and dependencies. Consider an input sequence $\{\mathbf{x}_0, \mathbf{x}_1, \cdots, \mathbf{x}_T\}$, where each $\mathbf{x}_t \in \mathbb{R}^{d_x}$. At each time step $t$, the RNN produces an output $\mathbf{y}_t \in \mathbb{R}^{d_y}$ and updates a hidden state $\mathbf{h}_t \in \mathbb{R}^{d_h}$. The key recurrence relations can be written as $\mathbf{h}_t = f(\mathbf{h}_{t-1}, \mathbf{x}_t)$ and $\mathbf{y}_t = g(\mathbf{h}_t)$
, where $f$ and $g$ are nonlinear maps, often implemented as parameterized neural layers. The hidden state $\mathbf{h}_t$ serves as a compact memory that carries information about the sequence’s history.

Unfolding the RNN over time, one can view it as a chain of repeating modules, each responsible for processing one element of the input sequence and updating the hidden state. This temporal unfolding makes it clear why RNNs can capture time-dependent patterns: the current hidden state is directly influenced by all previous inputs through a continuous update mechanism.

\subsubsection{Long Short-Term Memory}

While vanilla RNNs can model temporal dependencies, they often struggle to learn long-range correlations due to issues like vanishing or exploding gradients. To address this, the LSTM architecture introduces a cell state $\mathbf{c}_t \in \mathbb{R}^{d_h}$, which can propagate information over long time spans more stably.

An LSTM cell maintains both $\mathbf{h}_t$ and $\mathbf{c}_t$, and employs gating mechanisms to control information flow. At each time step $t$, given an input $\mathbf{x}_t$ and the previous states $\mathbf{h}_{t-1}$, $\mathbf{c}_{t-1}$, we first form a concatenated vector $\mathbf{v}_t = [\mathbf{h}_{t-1}; \mathbf{x}_t] \in \mathbb{R}^{d_x + d_h}$. The LSTM equations are:
\begin{subequations} \label{eq:lstm}
\begin{align}
\mathbf{f}_t &= \sigma(\mathbf{W}_f \mathbf{v}_t + \mathbf{b}_f), \\
\mathbf{i}_t &= \sigma(\mathbf{W}_i \mathbf{v}_t + \mathbf{b}_i), \\
\tilde{\mathbf{C}}_t &= \tanh(\mathbf{W}_C \mathbf{v}_t + \mathbf{b}_C), \\
\mathbf{c}_t &= \mathbf{f}_t \odot \mathbf{c}_{t-1} + \mathbf{i}_t \odot \tilde{\mathbf{C}}_t, \label{eq:lstm_state_updates} \\
\mathbf{o}_t &= \sigma(\mathbf{W}_o \mathbf{v}_t + \mathbf{b}_o), \\
\mathbf{h}_t &= \mathbf{o}_t \odot \tanh(\mathbf{c}_t).
\end{align}
\end{subequations}

Here, $\sigma(\cdot)$ is the sigmoid activation, $\tanh(\cdot)$ is the hyperbolic tangent, $\mathbf{W}_\alpha$ and $\mathbf{b}_\alpha$ are trainable parameters, and $\odot$ denotes element-wise multiplication. The input gate $\mathbf{i}_t$ and candidate state $\tilde{\mathbf{C}}_t$ determine what new information to add, the forget gate $\mathbf{f}_t$ determines what information to remove from the cell state, and the output gate $\mathbf{o}_t$ decides what filtered information is sent to the hidden state. This design allows gradients to flow more effectively through time, enabling the LSTM to capture complex temporal patterns over longer sequences.

\subsection{Variational Quantum Circuits for Sequence Modeling}
VQCs are quantum neural network models characterized by a set of trainable parameters (circuit angles) that are adjusted via classical optimization. Given an input vector $\mathbf{x} \in \mathbb{R}^{d}$, we first encode it into an $N$-qubit quantum state, $|\psi(\mathbf{x})\rangle = U_{\text{enc}}(\mathbf{x}) |0\rangle^{\otimes N}$
, where $U_{\text{enc}}(\mathbf{x})$ is a problem-specific encoding unitary. For instance, one may use single-qubit rotations $R_y(\theta_i)$ and $R_z(\phi_i)$, where $(\theta_i,\phi_i)$ are functions of the components of $\mathbf{x}$. A VQC then applies a parameterized unitary $U(\boldsymbol{\theta})$, composed of elementary gates such as rotations and CNOTs to transform the encoded quantum state into $|\phi(\mathbf{x}; \boldsymbol{\theta})\rangle = U(\boldsymbol{\theta}) U_{\text{enc}}(\mathbf{x}) |0\rangle^{\otimes N}$
, with $\boldsymbol{\theta}$ denoting the trainable parameters. Measurements in a chosen basis yield classical outputs $y_j(\mathbf{x}; \boldsymbol{\theta}) = \langle \phi(\mathbf{x}; \boldsymbol{\theta}) | M_j | \phi(\mathbf{x}; \boldsymbol{\theta}) \rangle$
, where $M_j$ is a measurement operator (e.g., $M_j = Z \otimes I \otimes \cdots$), producing a real-valued output. By stacking several such measurements or using multiple qubits, we obtain an output vector $\mathbf{y} \in \mathbb{R}^{d'}$ that can be further processed classically. VQCs can approximate nonlinear functions, enabling them to serve as building blocks for a variety of machine learning tasks. Their expressivity can, in principle, surpass that of classical neural networks under certain conditions, making them suitable candidates for encoding and processing temporal dependencies in sequence data.

\subsection{Formulating the QLSTM Cell}

To form a QLSTM, we replace the linear transformations (e.g., $\mathbf{W}_f[\mathbf{h}_{t-1}; \mathbf{x}_t] + \mathbf{b}_f$) with VQCs. Let $\mathbf{v}_t = [\mathbf{h}_{t-1}; \mathbf{x}_t] \in \mathbb{R}^{d_x + d_h}$ be the concatenation of the previous hidden state and the current input. In the QLSTM, the gates are computed as:
\begin{align}
f_t &= \sigma(\mathrm{VQC}_f(\mathbf{v}_t)), \\
i_t &= \sigma(\mathrm{VQC}_i(\mathbf{v}_t)), \\
\tilde{\mathbf{C}}_t &= \tanh(\mathrm{VQC}_C(\mathbf{v}_t)), \\
o_t &= \sigma(\mathrm{VQC}_o(\mathbf{v}_t)).
\end{align}
Each $\mathrm{VQC}_\alpha$ encodes $\mathbf{v}_t$ into a quantum state via the encoding unitary $U_{\text{enc}}(\mathbf{v}_t)$, applies $U(\boldsymbol{\theta}_\alpha)$, and measures the system to yield a vector in $\mathbb{R}^{d_h}$. Thus, the QLSTM cell state and hidden state updates remain the same as in Eq.~\eqref{eq:lstm_state_updates}, but the gating functions now leverage quantum operations.

\subsection{Partitioning for a Distributed QLSTM}
\begin{figure}[htbp]
\begin{center}
\includegraphics[width=1\columnwidth]{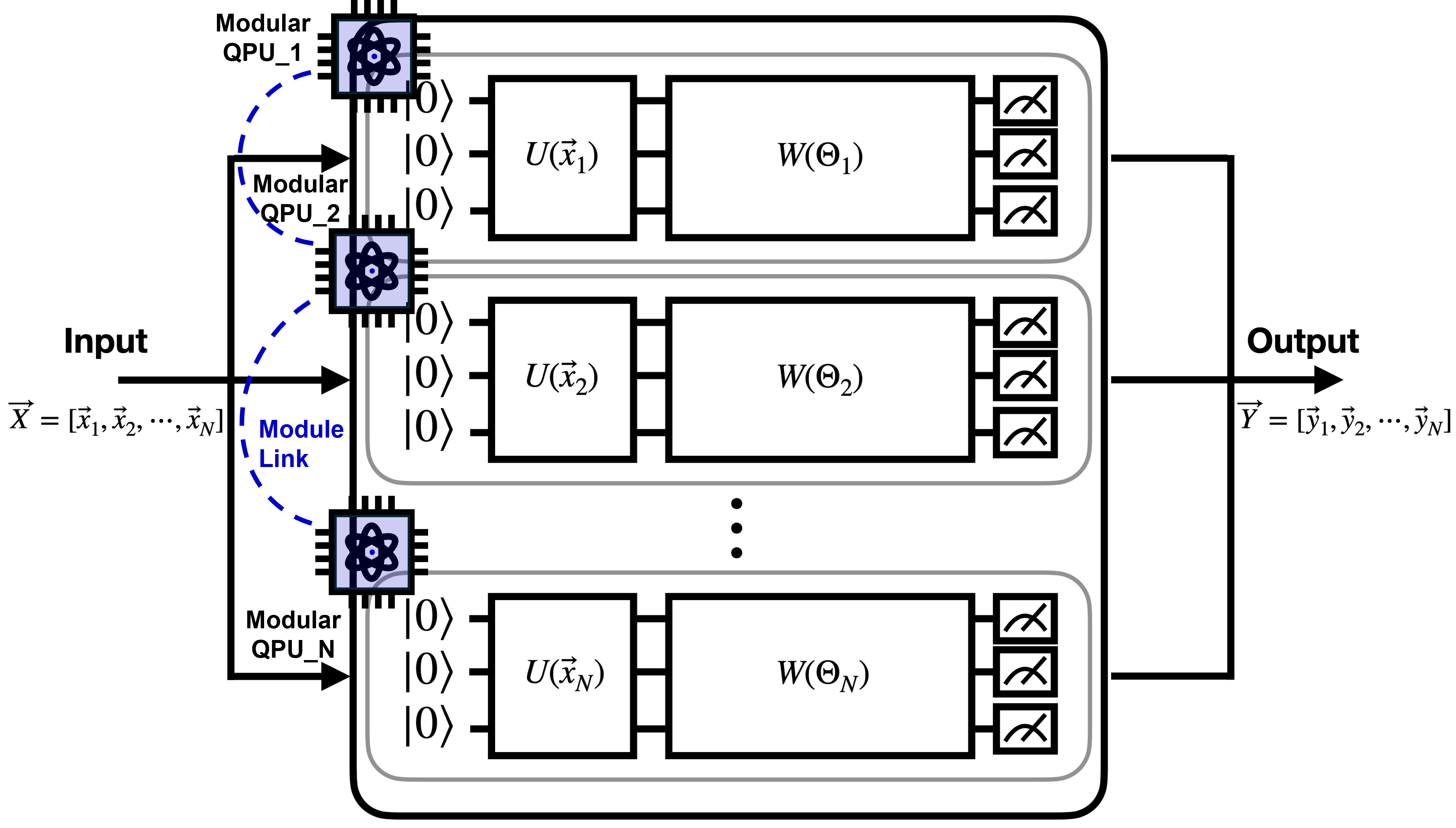}
\caption{{\bfseries Split Variational Quantum Circuits (VQCs) functioning as Quantum Neural Networks (QNN), as used in this paper.}}
\label{fig:split_VQC}
\end{center}
\end{figure}

\begin{figure}[htbp]
\begin{center}
\includegraphics[width=1\columnwidth]{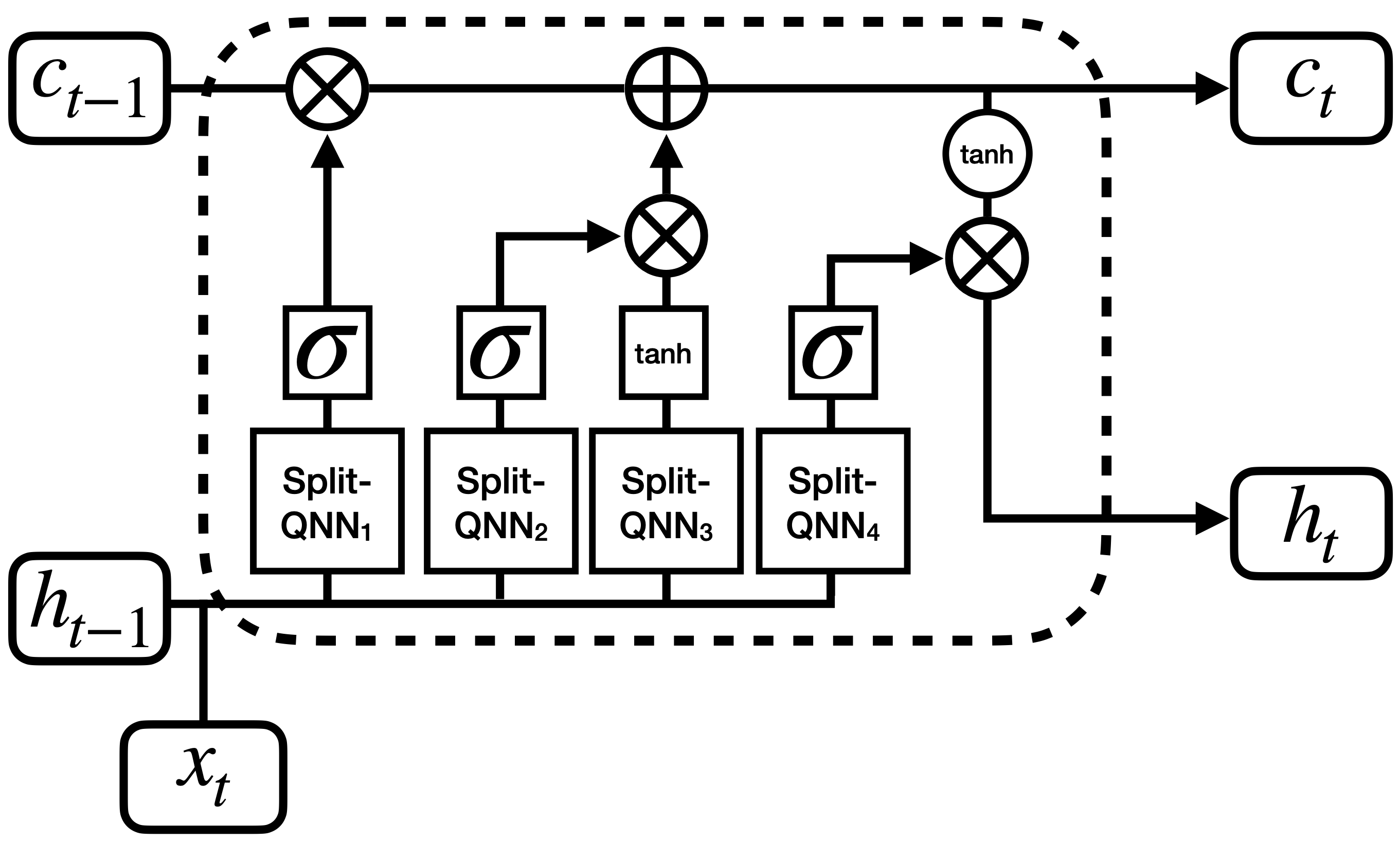}
\caption{{\bfseries Distributed QLSTM Framework via Split-QNN Cells.}}
\label{fig:split_QLSTM}
\end{center}
\end{figure}

As the dimension $D = d_x + d_h$ grows, implementing a single large VQC for each gate may become impractical. To address scalability, we partition $\mathbf{v}_t$ into $M$ sub-vectors, each handled by a smaller VQC (depicted in \figureautorefname{\ref{fig:split_VQC}}):
\begin{equation}
\mathbf{v}_t = [\mathbf{v}_t^{(1)}; \mathbf{v}_t^{(2)}; \cdots; \mathbf{v}_t^{(M)}],\quad \sum_{m=1}^{M} D_m = D.
\end{equation}
Each $\mathrm{VQC}_f^{(m)}$ is a smaller quantum circuit that encodes a subset of the input and output a vector with smaller dimensions, thus reducing circuit depth and complexity. These sub-VQCs are evaluated (potentially in parallel on separate QPUs) and the results are collected and concatenated.
For the forget gate, for example, the result can be expressed as:
\begin{equation}
f_t = \sigma([\mathrm{VQC}_f^{(1)}(\mathbf{v}_t^{(1)}), \cdots, \mathrm{VQC}_f^{(M)}(\mathbf{v}_t^{(M)})]) \in \mathbb{R}^{d_h},
\end{equation}
with $\sum_{m=1}^{M} d_h^{(m)} = d_h$. 
%
%
The same decomposition applies to $i_t$, $\tilde{\mathbf{C}}_t$, and $o_t$, as illustrated in \figureautorefname{\ref{fig:split_QLSTM}}. The cell and hidden state updates remain:
\begin{equation}
\mathbf{c}_t = f_t \odot \mathbf{c}_{t-1} + i_t \odot \tilde{\mathbf{C}}_t,\quad
\mathbf{h}_t = o_t \odot \tanh(\mathbf{c}_t),
\end{equation}
but now each gate’s computation is distributed. This modularization results in a distributed QLSTM, where multiple QPUs handle distinct subsets of the problem, thus enabling larger-scale quantum sequence models under NISQ-era constraints.
\subsection{Training and Optimization}

The QLSTM parameters $\{\boldsymbol{\theta}_f, \boldsymbol{\theta}_i, \boldsymbol{\theta}_C, \boldsymbol{\theta}_o\}$ are optimized via hybrid quantum-classical gradient-based methods. Gradients with respect to quantum circuit parameters are computed using the parameter-shift rule. For a generic expectation value $f(\mathbf{x}; \theta_i)$, the gradient can be computed as $\nabla_{\theta_i} f(\mathbf{x}; \theta_i) = \frac{1}{2} \left(f(\mathbf{x}; \theta_i + \frac{\pi}{2}) - f(\mathbf{x}; \theta_i - \frac{\pi}{2})\right)$.
%
By applying this rule to each trainable parameter across all VQCs (including those in the distributed setup), we can perform standard optimization techniques such as stochastic gradient descent or RMSProp. The result is an iterative procedure that updates $\boldsymbol{\theta}$ to minimize a loss function defined on a given temporal prediction or classification task.

\subsection{Resource Estimation for Distributed QLSTM}

In this subsection, we provide a resource estimation analysis for implementing a distributed, VQC-based QLSTM model. The goal is to understand how model complexity scales in terms of the number of trainable parameters, the number of qubits per VQC, and the total available quantum hardware resources (number of QPUs or quantum cores). By breaking down a large QLSTM into multiple smaller VQCs that run in parallel, we can potentially achieve greater overall capacity while remaining within the capabilities of near-term devices.

\subsubsection{Parameter Scaling in QLSTM}

A QLSTM cell comprises four primary gating functions: forget ($f_t$), input ($i_t$), candidate ($\tilde{\mathbf{C}}_t$), and output ($o_t$). In addition, there may be separate VQCs for transforming the internal cell state $\mathbf{c}_t$ into output $\mathbf{y}_t$ or new hidden states $\mathbf{h}_t$. Each gate takes as input a vector $\mathbf{v}_t = [\mathbf{h}_{t-1}; \mathbf{x}_t] \in \mathbb{R}^{D}$, where $D = d_x + d_h$. To manage large $D$, we partition it into $M$ segments:
\begin{equation}
\mathbf{v}_t = [\mathbf{v}_t^{(1)}; \mathbf{v}_t^{(2)}; \cdots; \mathbf{v}_t^{(M)}],
\end{equation}
with each $\mathbf{v}_t^{(m)} \in \mathbb{R}^{D_m}$ such that $\sum_{m=1}^{M} D_m = D$.

For each gate, we partition the input into $M$ smaller subspaces, each of dimension $D_{m}$. Rather than embedding all $D_{m}$ features via amplitude encoding (which would require $2^{q_{m}} \ge D_{m}$), we employ angle-based (or other local) encoding strategies that assign a small set of features directly to $q_{m}$ qubits. This approach does not demand $\log_{2}(D_{m})$ qubits. Instead, we typically choose $q_{m}$ to match or modestly exceed the local subspace dimension $D_{m}$ if we encode one or a few features per qubit. For simplicity, one may set $q_{m} = q$ across all partitions, leading to a total of $M q$ qubits. A more flexible approach allows $q_{m}$ to vary per partition to accommodate hardware constraints or specific data-encoding needs.

Each VQC is characterized by a set of tunable parameters, scaling with both the circuit depth $L$ and the number of qubits $q$. In a typical design with single-qubit rotations and controlled entangling gates, a VQC block of depth $L$ and $q$ qubits can carry 
\begin{equation}
    P_{\text{VQC}} \sim \mathcal{O}(L \cdot q),
\end{equation}
assuming each layer contributes a constant number of parameters per qubit. More complex parameterizations with multi-qubit rotations may scale as $\mathcal{O}(L \cdot q^{2})$. 

In a QLSTM cell, each of the four main gates ($f_{t}, i_{t}, \tilde{\mathbf{C}}_{t}, o_{t}$) employs $M$ of these sub-VQCs. Hence, the parameter count for these four gates is
\begin{equation}
    P_{\text{gates}} \approx 4 \cdot M \cdot P_{\text{VQC}}.
\end{equation}
If we include additional VQCs for output transformations or other processing, the total becomes
\begin{equation}
    P_{\text{total}} \approx (4 + N_{\text{extra}}) \cdot M \cdot P_{\text{VQC}},
\end{equation}
where $N_{\text{extra}}$ is the number of extra quantum blocks used (for instance, to map cell states or hidden states to final outputs). 

\subsubsection{Hardware Requirements and Parallelization}

To run $M$ sub-VQCs in parallel, we require $M$ distinct QPUs or a multi-core quantum processor architecture with at least $M$ quantum cores. Each core runs a VQC with $q$ qubits. Thus, the total qubit count across all QPUs is $M \times q$, but since each VQC runs independently, the hardware requirement per QPU remains moderate.

As $D$ grows (i.e., as we consider larger input and hidden dimensions), we can increase $M$ to keep $q$ relatively small. This allows the complexity on each individual QPU to remain within NISQ capabilities. The trade-off is an increase in classical overhead for coordinating multiple QPUs and aggregating their outputs.






\section{Results}
\label{Sec.III}

\subsection{Damped Harmonic Oscillator}

\begin{figure}[!t]
    \centering
    \includegraphics[width=1\columnwidth]{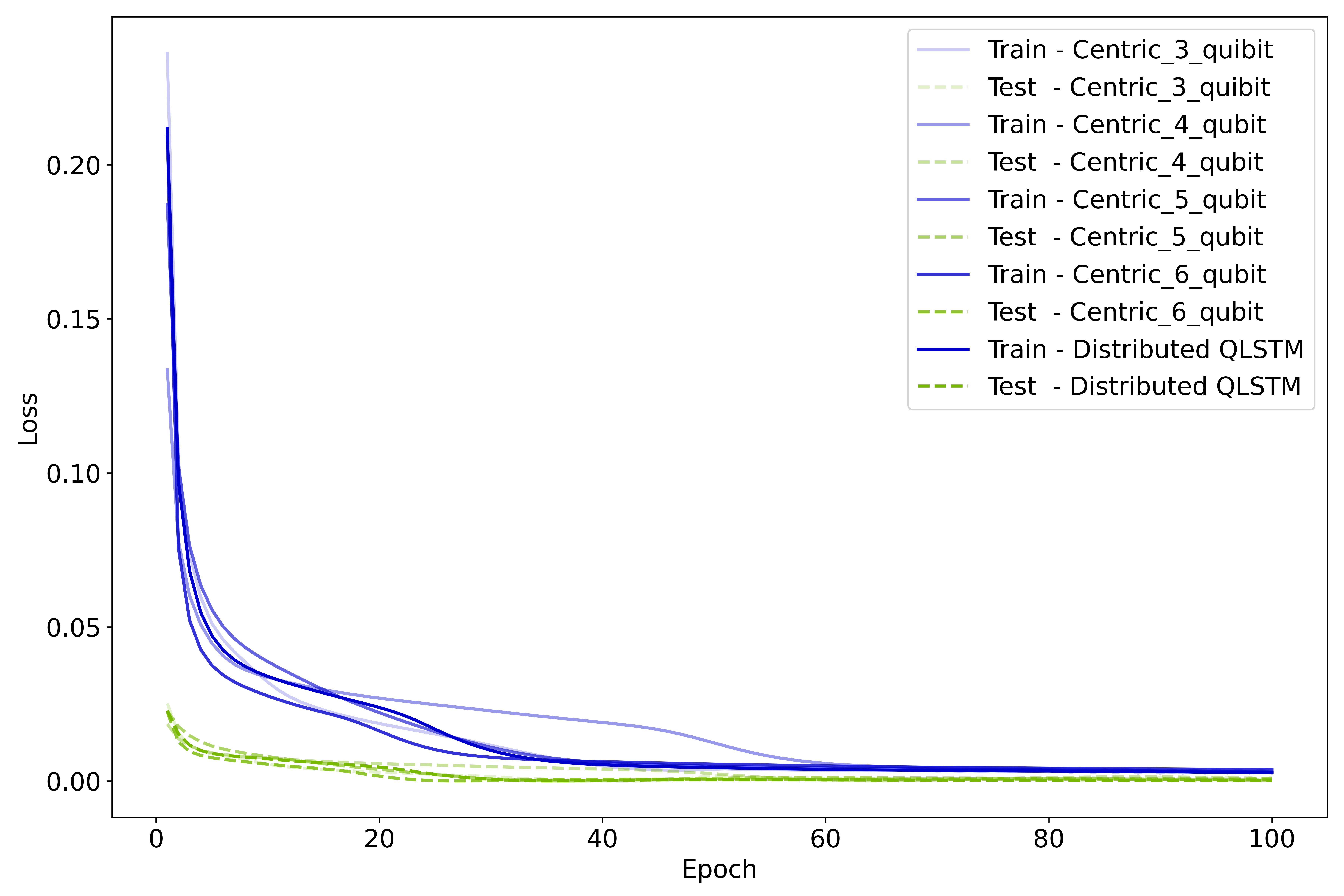}
    \caption{Training and testing loss curves comparing centric QLSTM models with different qubit counts (3, 4, 5, 6) and a distributed QLSTM approach, illustrating stable convergence for both methods over 100 epochs.}
    \label{fig:loss_plot}
\end{figure}

\begin{figure}[!t]
    \centering
    \includegraphics[width=1\columnwidth]{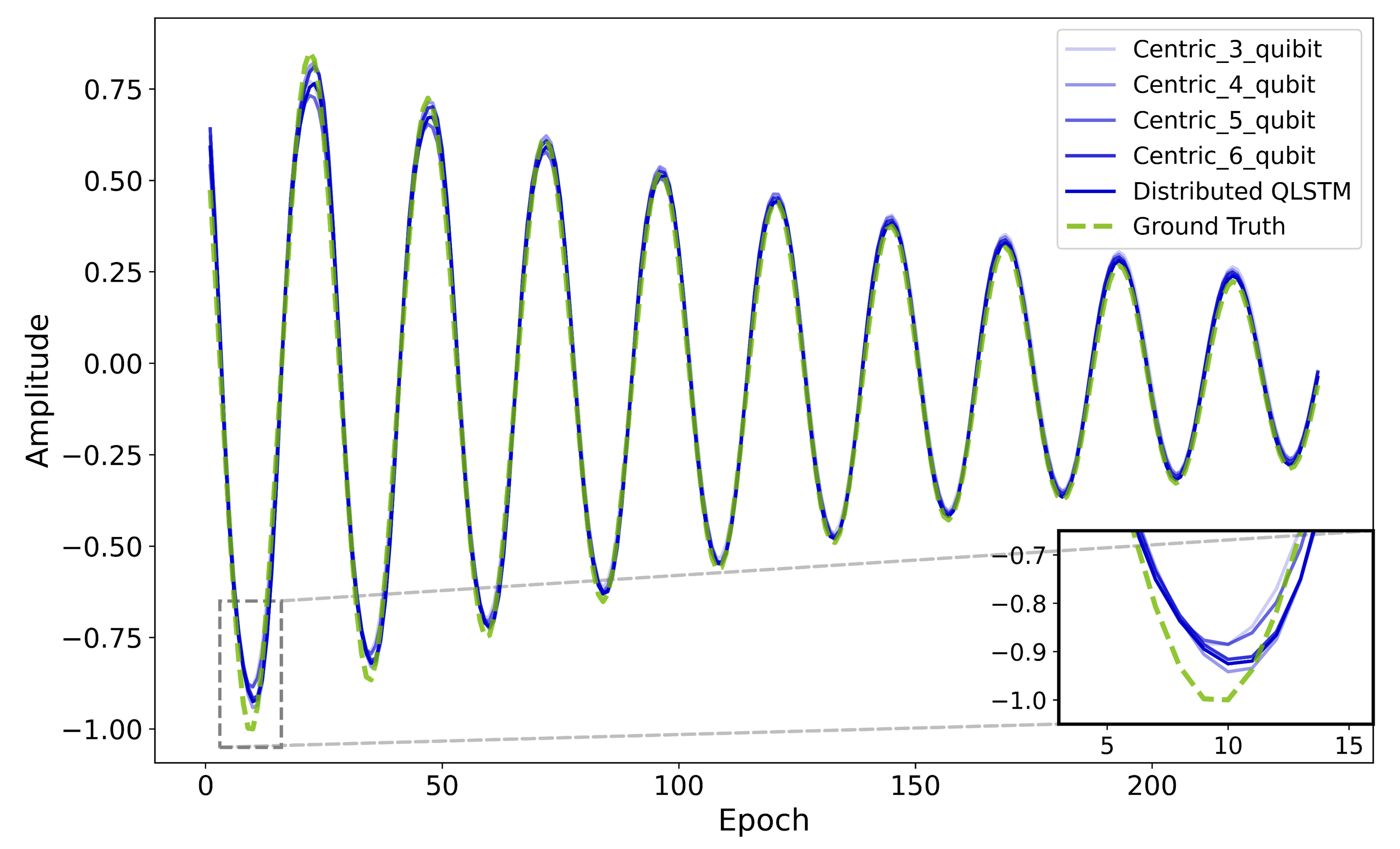}
    \caption{Time-series comparison of predicted angular velocities $\dot{\theta}(t)$ from the centric QLSTM (3, 4, 5, 6 qubits) and the distributed QLSTM relative to ground truth, demonstrating each model’s ability to capture both oscillatory and damped behavior.}
    \label{fig:time_series}
\end{figure}

To benchmark our distributed QLSTM, we first examine a damped harmonic oscillator, governed by
\begin{equation}
\frac{d^2 x}{dt^2} + 2\zeta \omega_0 \frac{dx}{dt} + \omega_0^2 x = 0,
\end{equation}
where $\omega_0=\sqrt{k/m}$ is the natural frequency and $\zeta$ is the damping ratio. We focus on a damped pendulum with parameters chosen to yield nontrivial oscillatory decay:
\begin{equation}
\frac{d^2 \theta}{dt^2} + \frac{b}{m}\frac{d\theta}{dt} + \frac{g}{L}\sin(\theta) = 0.
\end{equation}

We train the distributed QLSTM to predict the angular velocity $\dot{\theta}(t)$ given initial conditions $\theta(0)=0$ and $\dot{\theta}(0)=3\,\mathrm{rad/s}$. By partitioning the variational quantum circuits, the model can exploit parallelism and effectively learn the nonlinear behavior. \figureautorefname{\ref{fig:loss_plot}} shows stable convergence over 100 epochs, and \figureautorefname{\ref{fig:time_series}} verifies accurate tracking of the damped dynamics. These results (summarized in Table~\ref{tab:damped_oscillation}) demonstrate the feasibility of distributed quantum-classical architectures for complex time-series modeling. Further investigations will address scalability, noise mitigation, and broader applications of this approach in quantum machine learning.

\begin{table}[!t]
    \centering
    \caption{Performance metrics for the Damped Oscillation task using Centric QLSTM models with varying qubit numbers and the Distributed QLSTM model.}
    \begin{tabular}{@{}lcc@{}}
        Model & R-square & Testing Epoch Convergence \\
        \midrule
        Centric QLSTM (3 qubit) & 0.9898 & 81 \\
        Centric QLSTM (4 qubit) & 0.9913 & 68 \\
        Centric QLSTM (5 qubit) & 0.9938 & 38 \\
        Centric QLSTM (6 qubit) & 0.9903 & 27 \\
        Distributed QLSTM & 0.9936 & 31 \\
    \end{tabular}
    \label{tab:damped_oscillation}
\end{table}

\subsection{Nonlinear Autoregressive Moving Average Sequences}

\begin{figure}[!b]
    \centering
    \includegraphics[width=1\columnwidth]{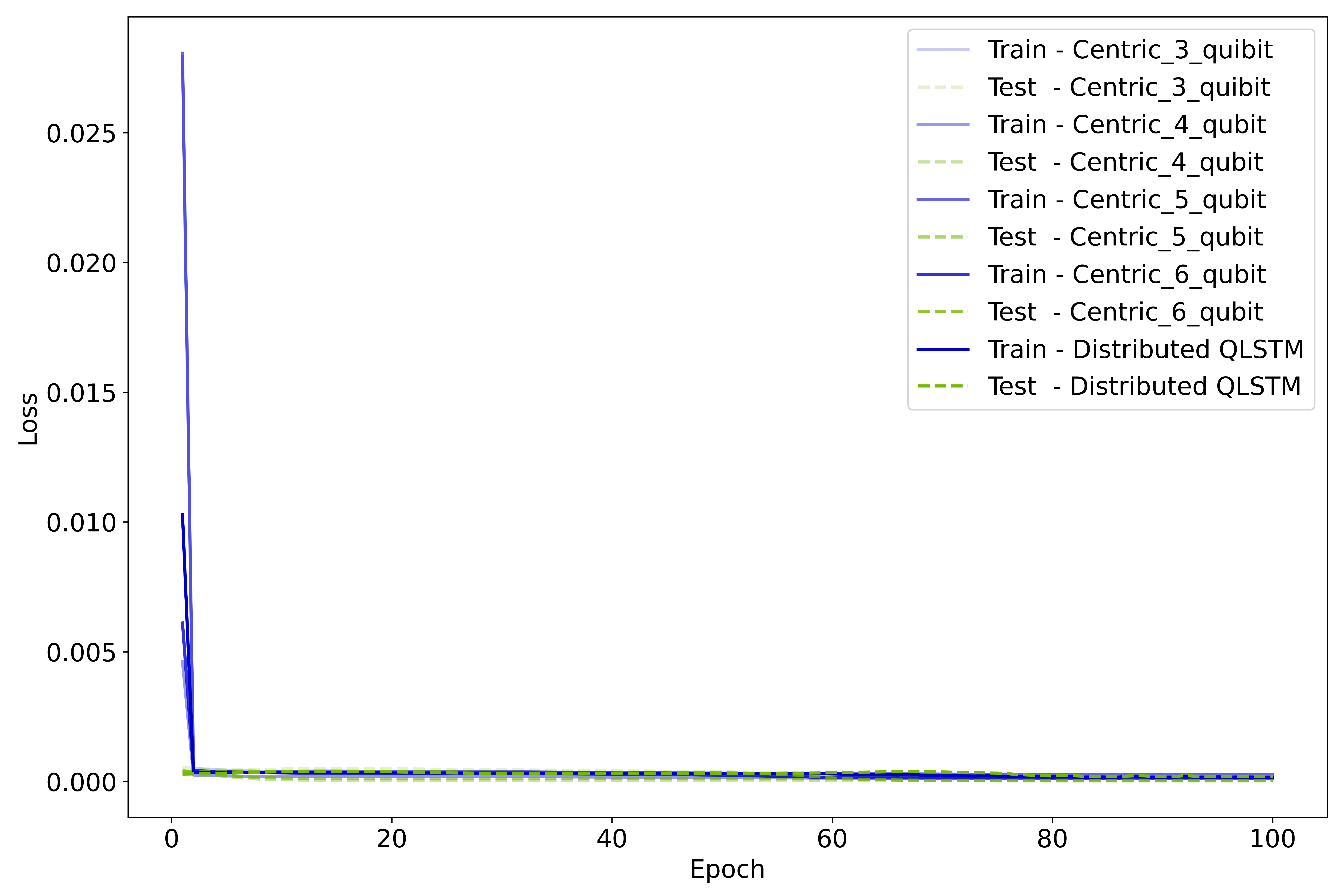}
    \caption{Training and testing loss curves for the centric QLSTM (3, 4, 5, 6 qubits) and the distributed QLSTM, demonstrating rapid convergence to near-zero loss for both approaches.}
    \label{fig:narma_loss}
\end{figure}

\begin{figure}[!b]
    \centering
    \includegraphics[width=1\columnwidth]{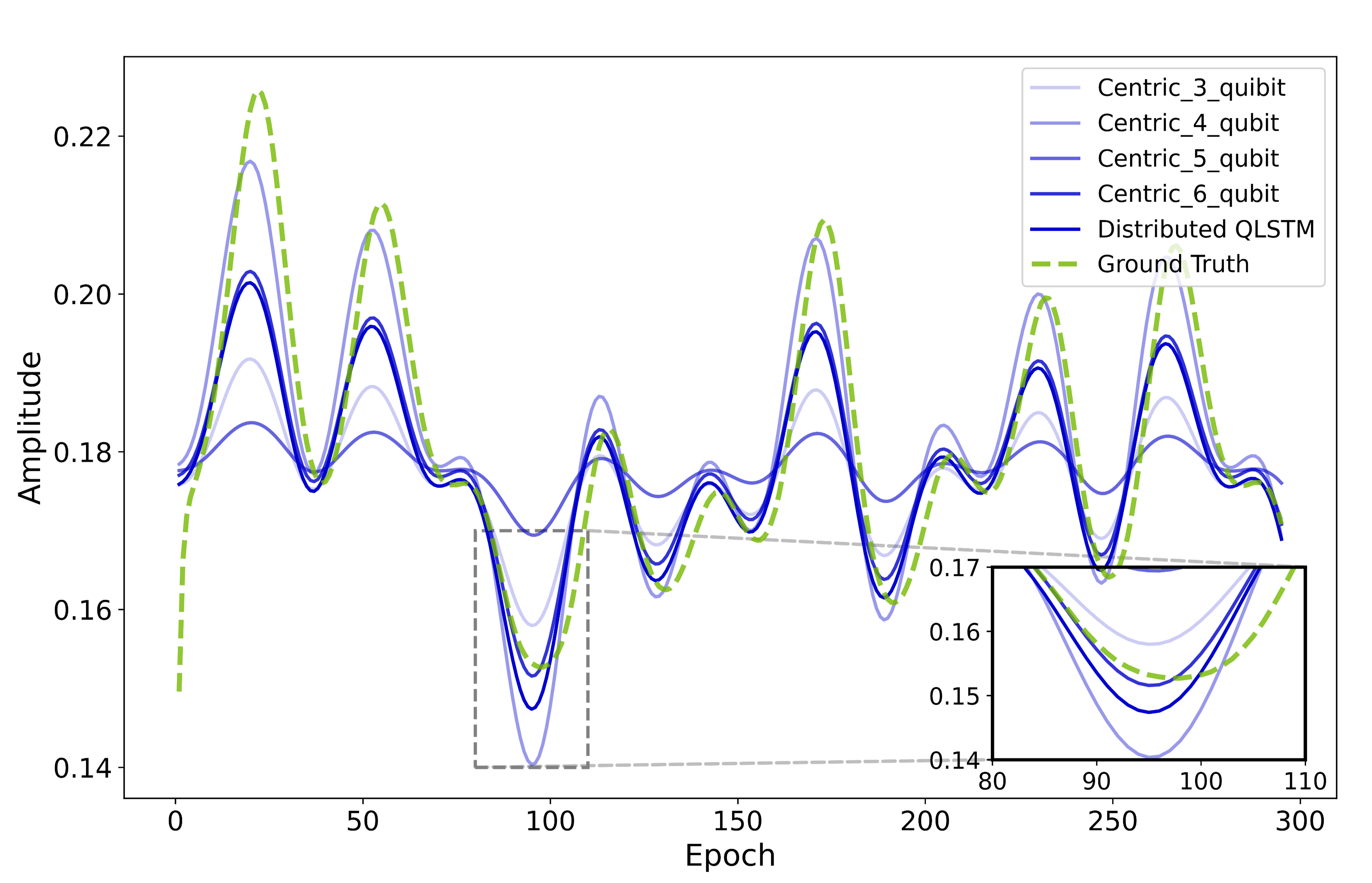}
    \caption{Time-series comparison of predicted amplitudes from the centric QLSTM (3, 4, 5, 6 qubits) and the distributed QLSTM against the ground truth, highlighting accurate modeling of the system’s oscillatory behavior across different quantum resource configurations.}
    \label{fig:narma_time_series}
\end{figure}

To further benchmark the distributed QLSTM framework, we employ the Nonlinear Autoregressive Moving Average (NARMA) family of sequences, a standard test for modeling complex, memory-intensive processes. A representative example is the NARMA-2 relation
\begin{equation}
y_{t+1} = 0.4 \, y_t + 0.4 \, y_t y_{t-1} + 0.6 \, u_t^3 + 0.1,
\end{equation}
where \(u_t\) is an external input. Higher-order variants (e.g., NARMA-5, NARMA-10) incorporate longer temporal histories, posing additional challenges for models to retain and process extended dependencies.

Data generation proceeds by creating a rich input stream \(\{u_t\}\) (e.g., via random or trigonometric transformations) and applying the chosen NARMA-\(n\) equation to obtain the target outputs \(\{y_t\}\). Figures~\ref{fig:narma_loss} and \ref{fig:narma_time_series} confirm that the distributed QLSTM converges reliably and accurately reproduces the nonlinear, autoregressive structure of the NARMA sequences. Similarly, these findings (summarized in Table~\ref{tab:narma}) suggest that the partitioned quantum-classical architecture can tackle complex temporal tasks and may scale effectively to higher-order prediction challenges. Further exploration is warranted to clarify potential quantum advantages in broader time-series regimes.

\begin{table}[!t]
    \centering
    \caption{Performance metrics for the NARMA task using Centric QLSTM models with varying qubit numbers and the Distributed QLSTM model.}
    \begin{tabular}{@{}lcc@{}}
        Model & R-square & Testing Epoch Convergence \\
        \midrule
        Centric QLSTM (3 qubit) & 0.5477 & 2 \\
        Centric QLSTM (4 qubit) & 0.8611 & 2 \\
        Centric QLSTM (5 qubit) & 0.2651 & 2 \\
        Centric QLSTM (6 qubit) & 0.7799 & 2 \\
        Distributed QLSTM & 0.7523 & 2 \\
    \end{tabular}
    \label{tab:narma}
\end{table}

\section{\uppercase{Conclusions}}
\label{Sec.IV}

This study introduces a scalable framework for Distributed QLSTM, addressing the limitations of classical methods in learning complex temporal dependencies. By leveraging modular quantum computing, the framework distributes computational tasks across multiple QPUs, optimizing resource utilization while maintaining performance. The results on damped harmonic oscillator and NARMA benchmarks demonstrate that the distributed QLSTM achieves stable training and accurate temporal predictions, validating its effectiveness in handling nonlinear, memory-intensive systems.

The modular architecture proposed in this work lays the foundation for broader adoption of distributed quantum-classical systems in machine learning and quantum information processing. Future efforts will focus on refining partitioning strategies, mitigating quantum noise, and exploring applications in more complex quantum systems. This research underscores the transformative potential of distributed quantum machine learning in addressing scalability and complexity challenges, paving the way for its integration into advanced computational paradigms.

\section*{Acknowledgment}
This work was supported by the Engineering and Physical Sciences Research Council (EPSRC) under grant number EP/W032643/1.

\clearpage
\bibliographystyle{ieeetr}
{\small
\bibliography{example,bib/classical_ml,bib/qml_examples,bib/vqc,bib/nisq}}

\vspace{12pt}

\end{document}